\documentclass[apjl]{emulateapj}

\usepackage{amsmath}
\usepackage{graphicx}
\usepackage{epsfig,psfrag,epic,eepic}
\usepackage{textcomp}
\usepackage{times}

\slugcomment{Accepted for Publication in ApJ Letters, May 12, 2010}

\shorttitle{Imaging of LkCa~15 Disk Gap in Reflected Light}
\shortauthors{Thalmann et al.}

\begin{document}

\title{Imaging of a Transitional Disk Gap in Reflected Light:\\ Indications of Planet Formation Around the Young Solar Analog L\lowercase{k}C\lowercase{a}~15\altaffilmark{$\star$}}

\author{C. Thalmann\altaffilmark{1}, C.~A. Grady\altaffilmark{2},
	M. Goto\altaffilmark{1}, J.~P. Wisniewski\altaffilmark{3},
	M. Janson\altaffilmark{4}, T. Henning\altaffilmark{1},  
	M. Fukagawa\altaffilmark{5}, M. Honda\altaffilmark{6}, 
	G.~D. Mulders\altaffilmark{7,8},
	M. Min\altaffilmark{9},
	A. Moro-Mart\'in\altaffilmark{10},
	M.~W. McElwain\altaffilmark{11},
	K.~W. Hodapp\altaffilmark{12},
	J. Carson\altaffilmark{1,13},
	L. Abe\altaffilmark{14},
	W. Brandner\altaffilmark{1},
	S. Egner\altaffilmark{15},
	M. Feldt\altaffilmark{1},
	T. Fukue\altaffilmark{16},
	T. Golota\altaffilmark{15},
	O. Guyon\altaffilmark{15},
	J. Hashimoto\altaffilmark{16},
	Y. Hayano\altaffilmark{15},
	M. Hayashi\altaffilmark{15}, 
	S. Hayashi\altaffilmark{15}, 
	M. Ishii\altaffilmark{15},
	R. Kandori\altaffilmark{16}, 
	G.~R. Knapp\altaffilmark{11},
	T. Kudo\altaffilmark{16},  
	N. Kusakabe\altaffilmark{16},
	M. Kuzuhara\altaffilmark{16,17}, 
	T. Matsuo\altaffilmark{16},
	S. Miyama\altaffilmark{16}, 
	J.-I. Morino\altaffilmark{16},
	T. Nishimura\altaffilmark{15},
	T.-S. Pyo\altaffilmark{15},
	E. Serabyn\altaffilmark{18}, 
	H. Shibai\altaffilmark{5},
	H. Suto\altaffilmark{16},
	R. Suzuki\altaffilmark{16}, 
	M. Takami\altaffilmark{19},
	N. Takato\altaffilmark{15},
	H. Terada\altaffilmark{15},
	D. Tomono\altaffilmark{15},
	E.~L. Turner\altaffilmark{11,20},
	M. Watanabe\altaffilmark{21},
	T. Yamada\altaffilmark{22},
	H. Takami\altaffilmark{15},
	T. Usuda\altaffilmark{15},
	M. Tamura\altaffilmark{16}
}

\altaffiltext{$\star$}{Based on data collected at Subaru Telescope, which
	is operated by the National Astronomical Observatory of Japan, and on
	data obtained from the ESO Science Archive Facility under program ID
	280.C-5033(A).}
\altaffiltext{1}{Max Planck Institute for Astronomy, Heidelberg, Germany;
	\texttt{thalmann@mpia.de}.}
\altaffiltext{2}{Eureka Scientific and Goddard Space Flight Center, Greenbelt, USA.}
\altaffiltext{3}{University of Washington, Seattle, Washington, USA.}
\altaffiltext{4}{University of Toronto, Toronto, Canada.}
\altaffiltext{5}{Osaka University, Osaka, Japan.}
\altaffiltext{6}{Faculty of Science, Kanagawa University, Kanagawa, Japan.}
\altaffiltext{7}{Astronomical Institute ``Anton Pannekoek'', University of Amsterdam,
	Amsterdam, The Netherlands.}
\altaffiltext{8}{SRON Netherlands Institute for Space Research, Groningen, The
	Netherlands.}
\altaffiltext{9}{Astronomical Inst.,\ Univ.\ of Utrecht, Utrecht,
	The Netherlands.}
\altaffiltext{10}{Dep.\ of Astrophysics, CAB - CSIC/INTA, Madrid, Spain.}
\altaffiltext{11}{Dep.\ of Astrophysical Sciences, Princeton Univ.,\ Princeton, USA.}
\altaffiltext{12}{Institute for Astronomy, University of Hawai`i, Hilo, Hawai`i, USA.}
\altaffiltext{13}{College of Charleston, Charleston, South Carolina, USA.}
\altaffiltext{14}{Laboratoire Hippolyte Fizeau, Nice, France.}
\altaffiltext{15}{Subaru Telescope, Hilo, Hawai`i, USA.}
\altaffiltext{16}{National Astronomical Observatory of Japan, Tokyo, Japan}
\altaffiltext{17}{University of Tokyo, Tokyo, Japan.}
\altaffiltext{18}{JPL, California Institute of Technology, Pasadena, USA.}
\altaffiltext{19}{Inst.\ of Astronomy and Astrophysics, Academia Sinica, Taipei, Taiwan.}
\altaffiltext{20}{Institute for the Physics and Mathematics of the Universe, University
	of Tokyo, Japan.}
\altaffiltext{21}{Department of Cosmosciences, Hokkaido University, Sapporo, Japan.}
\altaffiltext{22}{Astronomical Institute, Tohoku University, Sendai, Japan}

\begin{abstract}\noindent
We present $H$- and $K_\mathrm{s}$-band imaging data resolving 
the gap in the transitional disk around LkCa~15, revealing the 
surrounding nebulosity.  We detect sharp 
elliptical contours delimiting the nebulosity on the inside as well as 
the outside, consistent with the shape, size, ellipticity, and
orientation of starlight reflected from the far-side disk wall, whereas 
the near-side wall is shielded from view by the disk's optically thick 
bulk.  We note that forward-scattering of starlight on the near-side 
disk surface could provide an alternate interpretation of the nebulosity.
In either case, this discovery provides confirmation of the disk geometry
that has been proposed to explain the spectral energy distributions (SED)
of such systems, comprising an optically thick disk with an 
inner truncation radius of $\sim$46\,AU enclosing a largely evacuated gap.
Our data show an offset of the nebulosity contours along the major axis,
likely corresponding to a physical pericenter offset of the disk gap.
This reinforces the leading theory that dynamical clearing by at least one
orbiting body is the cause of the gap.  Based on evolutionary models, our 
high-contrast imagery imposes an upper limit of 21\,$M_\mathrm{Jup}$ on 
companions at separations outside of 0\farcs1 and of 13\,$M_\mathrm{Jup}$ 
outside of 0\farcs2.  Thus, we find that a planetary system around LkCa~15
is the most likely explanation for the disk architecture.
\end{abstract}





\section{Introduction}


The circumstellar disks of gas and dust around newly formed stars are believed to
be the birthplaces of giant planets.
In some protoplanetary disks, evidence of 
gaps or inner cavities has been revealed through analysis of the infrared spectral
energy distribution \citep[SED, e.g.][]{calvet2002} or interferometry 
at infrared \citep[e.g.][]{ratzka2007} or millimeter wavelengths \citep[e.g.][]{brown2009}.  These objects have been termed ``transitional disks'', since they are thought to represent a transitional state of partial disk dissipation between the protoplanetary disk stage and the debris disk stage, with a variety of mechanisms proposed to account for central cavities: Photoevaporation, disk instabilities, and dynamical clearing \citep{alexander2006,chiang2007,bryden1999}.

One such system is \object{LkCa 15} (K5:Ve, V=11.91\,mag, J=9.42\,mag, H=8.6\,mag, K=8.16\,mag), a 0.97\,$M_\odot$, 
3--5\,Myr old \citep{simon2000},
weak-line T~Tauri star with a gas-rich millimeter-bright disk \citep{pietu2007,henning2010}. Recently, it was identified as a pre-transitional disk \citep{espaillat2007}, which consists of an inner disk component near the dust sublimation radius (0.12--0.15\,AU), a wide gap, and an outer, disk beyond 46 AU \citep{espaillat2008}.  Spitzer IRS spectra by \citet{sargent2009} reveal a relatively low level
of dust grain growth.
With an expected angular scale of 0\farcs33, the outer wall is 
accessible to 8-m class telescopes equipped with high-contrast imagers. 

\section{Observations and data reduction}

\begin{figure}[t]
\centering
\vspace{2mm}
\includegraphics[width=0.95\linewidth]{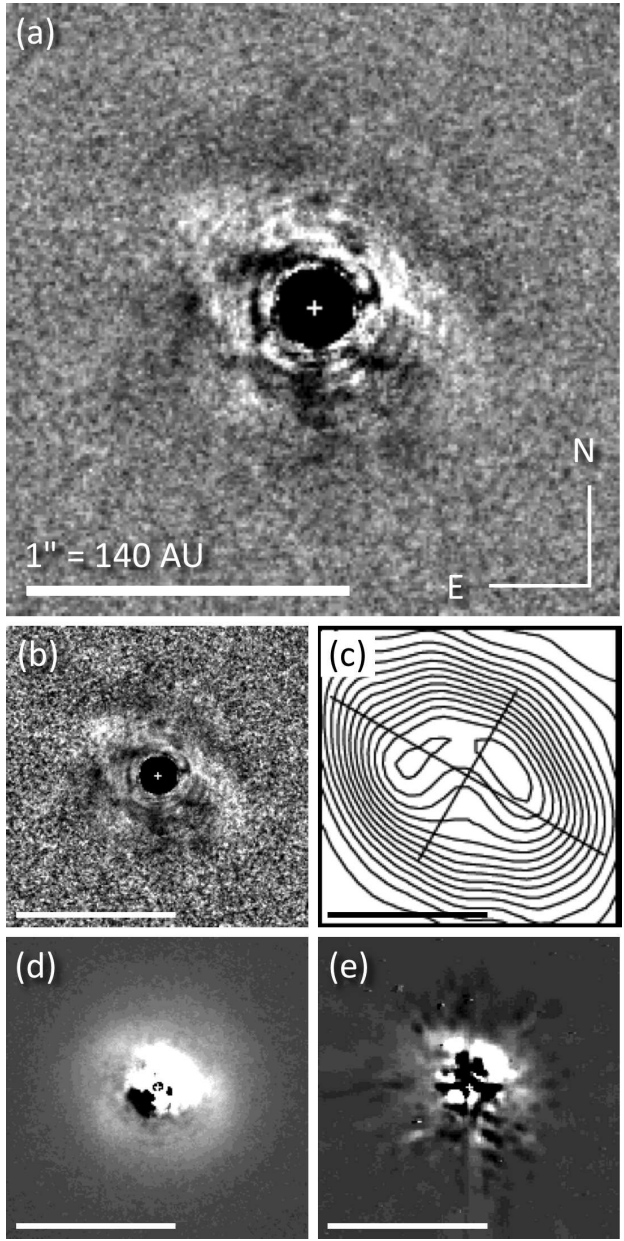}
\vspace{2mm}
\caption{High-contrast imaging of LkCa~15.  The scale
	bars in all panels correspond to 1$\arcsec$ = 140\,AU in
	projected separation. The greyscale is linear.
	(a) HiCIAO $H$-band image after conservative LOCI ADI.
	The inner edge of the outer disk is clearly 
	visible. The blacked-out area in the center represents the range of 
	separations where the field rotation is insufficient for ADI.  The
	negative features within 0\farcs5 are areas oversubtracted by LOCI
	as a reaction to the nearby positive features.  The brightest part
	of the nebulosity corresponds to $H=12.7$\,mag/arcsec$^2$.  
	(b) The signal-to-noise map of the same image, calculated
	in concentric annuli around the star.  The outer wings of the 
	nebulosity are rendered more visible.  The stretch is 
	$[-2.5\,\sigma,2.5\,\sigma]$.  
	(c) Interferometry image of the disk
	gap at 1.4\,mm taken from \citet{pietu2006}. The cross represents
	the major and minor axes of the nebulosity.
	(d) The HiCIAO $H$-band image 
	after reference PSF subtraction. The circular halo
	out to $\sim$0\farcs6 is most likely an artifact from imperfect
	PSF matching. 
	(e) The
	NACO $K_\mathrm{s}$-band image after reference PSF subtraction,
	for comparison.
}
\label{f:images}
\end{figure}

\subsection{Subaru/HiCIAO data}

As part of the ongoing high-contrast imaging survey SEEDS 
\citep[Strategic Exploration of Exoplanets and Disks with 
Subaru/\-HiCIAO;][]{tamura2009}, we observed LkCa~15 on December 26, 
2009, with the HiCIAO instrument \citep{hodapp2008}
on the Subaru Telescope with a field
of view of 20\arcsec$\times$20\arcsec\ and a plate scale of 9.5\,mas/pixel.
The images were 
taken in $H$ band (1.6\,\textmu m), and the image rotator
was operated in pupil-tracking mode to enable angular differential 
imaging \citep[ADI,][]{marois2006}.  The data set comprises 168 
frames, each containing the co-add of
3 unsaturated exposures of 1.39\,s, for a total integration time of 
700.6\,s. 
The image resolution provided by the adaptive optics system AO188
was close to 
the diffraction limit, with a full width at half maximum (FWHM) of 
55\,mas at 0\farcs6 natural seeing in $H$.

The data were corrected for flatfield and field distortion.
In order to search for point sources such as planets, ADI was then 
performed on the
images using the LOCI algorithm \citep[Locally Optimized Combination
of Images,][]{lafreniere2007}.  This form of ADI
is the most powerful high-contrast
imaging method currently available, as evidenced by recent direct
detections of substellar companions 
\citep{marois2008,thalmann2009}.
 
For two frames to be eligible for mutual point-spread function (PSF) 
subtraction at a given working radius, we require the differential 
field rotation arc at that radius to be at least 3\,FWHM = 165\,mas,
as opposed to the 0.75\,FWHM typically used for the detection of 
point-sources.  This protects extended structures of up to the given 
azimuthal size scale from self-subtraction.  We refer to this 
technique as ``conservative LOCI''. 
A circularly symmetric circumstellar structure is still eliminated 
under these conditions, but as our data demonstrate, a
sufficiently elliptical structure can survive by intersecting
the concentric annuli around the star at a steep angle.


An alternate method to reveal circumstellar nebulosity is subtraction 
of the PSF of a reference star, which avoids disk self-subtraction at
the price of less effective stellar PSF removal.  We observed the star 
\object{SAO 76662} in pupil-tracking mode immediately after LkCa~15 
for this purpose.  A neutral density filter was used with the 
adaptive optics wave-front sensor to achieve the same PSF quality as
with the fainter LkCa~15.  The exposures were centered, flatfielded, 
distortion-corrected, and directly collapsed with a pixel-wise median.
We subtracted the resulting reference PSF from each of the LkCa~15
frames before de-rotation and co-addition to ensure optimal matching
of the pupil-stabilized stellar PSFs.

\subsection{Archival VLT/NACO data}

For comparison, we retrieved $K_\mathrm{s}$-band coronagraphic 
imaging data of LkCa~15 taken
with VLT/NACO in 2007 from the ESO Science Archive Facility under the
program ID 280.C-5033(A).  The observations were made in field-tracking
mode and thus could not be used for LOCI.  However, a reference
star is included in the data set, allowing PSF subtraction.  The 
camera with 13\,mas pixel scale and the four-quadrant phase mask 
coronagraph 
had been used for this run, producing high contrast at a small 
inner working angle. 

\section{Results}

\subsection{Imaging of the disk gap}

The $H$-band HiCIAO images reveal a crescent-shaped nebulosity around 
LkCa 15 after both conservative LOCI reduction with a frame selection
criterion of 3\,FWHM and conventional reference PSF 
subtraction (Figure~\ref{f:images}). 
With a diameter of $\sim$$1\farcs2$ ($\approx 170$\,AU), it is 
elongated along the position angle $\sim$$60^\circ$, leaving
localized traces in the concentric annuli around the star in which LOCI
operates.  The consistency of the LOCI and reference subtraction images
demonstrates that the nebulosity is elliptical enough to survive 
conservative LOCI largely intact.  Although some flux is 
inevitably lost in this process, the elimination of the star's PSF in
LOCI is superior to
conventional reference PSF subtraction, revealing the sharp inner and 
outer edges of the structure clearly.  
The crescent is also found in the PSF-subtracted $K_\mathrm{s}$-band
images from VLT/NACO, supporting the interpretation that it represents
a real astronomical feature.  The gap enclosed by the inner
edge matches the gap predicted from the SED \citep{espaillat2007}
as well as the localized flux deficit seen in 1.4\,mm
interferometry \citep[cf.\ Figure~\ref{f:images}]{pietu2006} in terms
of size, ellipticity, and position angle.

\begin{figure}[t!]
\centering
\includegraphics[width=\linewidth, trim=2mm 0mm 2mm 0mm]{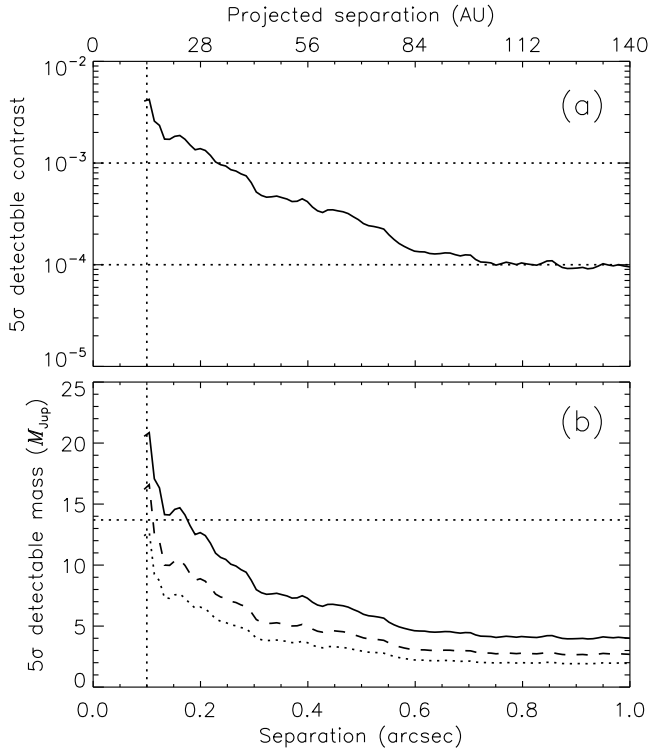}
\caption{Contrast (a) and mass (b) of companions detectable at 
	$5\,\sigma$ around LkCa~15, based on the $H$-band LOCI image after 
    convolution with a circular aperture of 4 pixel (38\,mas) in
    diameter.  The
    contrast values are converted into companion mass using the 
    \texttt{COND} evolutionary models by \citet{baraffe2003} assuming
    a distance of 140\,pc and an age of 5\,Myr (solid curve), 
    3\,Myr (dashed curve), and 1.5\,Myr (dotted curve), respectively.
	The vertical dotted lines mark the inner working angle of ADI in
	this data set.  The horizontal dotted lines guide the eye in (a)
	and indicate the deuterium burning limit of 13.7\,$M_\mathrm{Jup}$
	in (b).
	No companion candidates are detected.}
\label{f:planets}
\vspace*{1mm}
\end{figure}

The negative areas in the inner field represent oversubtraction,
which is inevitable in LOCI when strong signals are present.  The 
algorithm attempts minimize the root of mean squares of the residual
image, thus the positive mean offset of a signal will be lost, given
that it cannot be distinguished from the positive mean offset of the
stellar PSF halo.  In conservative LOCI, the frame selection criterion
that allows the survival of azimuthally extended positive flux 
concentrations also produce extended oversubtraction areas.
This effect renders the flux levels in the resulting image unreliable,
but preserves sharp edges in the original flux distribution. Similarly,
the location of a point source in regular LOCI is reliable, but its 
photometry is not, requiring the determination of a correction factor.

In its brightest pixel, the crescent in the LOCI image reaches 
a peak intensity of $H=12.7$\,mag/arcsec$^2$, which is to be taken as a 
conservative lower limit.  Reference PSF subtraction is expected to 
retain all of the disk flux, but the central part of the image is 
dominated by strong residuals from the PSF subtraction and therefore
does not provide a better photometric measurement.

The small extended signal $0.35\arcsec$ to the southeast
of the star in the $H$-band images likely corresponds to a real 
physical feature, given that it is visible both after LOCI processing
and after reference PSF subtraction (Figures~\ref{f:images}a, 
\ref{f:images}d).

\subsection{Constraints on point sources}

\subsection{Pericenter offset}

\begin{figure}[t!]
\centering
\includegraphics[width=\linewidth, trim=2mm 0mm 2mm 0mm]{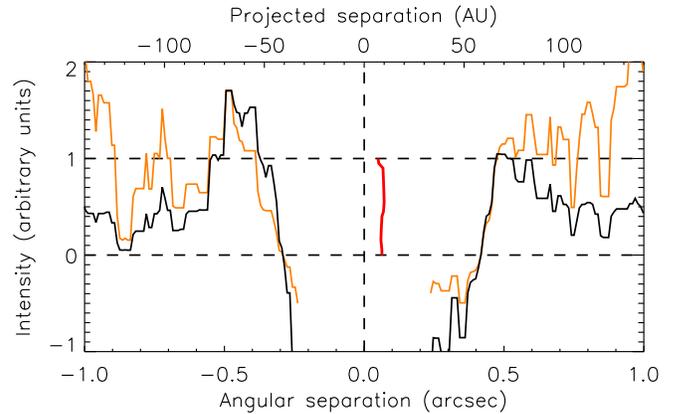}
\caption{Profile of the disk wall edges derived from an 
	11 pixel wide strip cut from the LOCI image along the perceived
	major axis of the ellipses in Figure~\ref{f:ellipses} (position
	angle $-32.8^\circ$), collapsed along the strip width and then
	median-smoothed on the scale of 5 pixels ($\sim$1\,FWHM), shown
	as a black curve.  Median smoothing
	conserves edges well while removing pixel-to-pixel noise.  
	The red	vertical curve in the center is the bisector of the two 
	opposed slopes of the profile, demonstrating the asymmetry. 
	The orange curve shows the same smoothed profile taken from the 
	LOCI image after rescaling with a map of the squared distance
	$r^2$ from the star.
	The dashed horizontal lines
	guide the eye.
}
\vspace*{2mm}
\label{f:bisector}
\end{figure}

Our LOCI imaging allows us to set upper limits on the point sources
present in the vicinity of LkCa~15.  Figure~\ref{f:planets} shows the
companion mass we can exclude at the $5\,\sigma$ level
as a function of separation, assuming the \texttt{COND} evolutionary
models by \citet{baraffe2003}, a distance of 140\,pc, and an age
of 3--5\,Myr \citet{simon2000}.  Since LkCa~15 is part of the Taurus
star-forming region, for which \citet{watson2009} give an age of
1--2\,Myr, we also plot a detectable mass curve for the age of 
1.5\,Myr.  The image was convolved
with a circular aperture 4 pixel in diameter, and the noise level is
calculated 
as the standard deviation in concentric annuli.  We compensate for the
expected flux loss due to partial self-subtraction by implanting test 
point sources in the raw frames and measuring how they are affected by 
the LOCI algorithm. 
We note that the bright nebulosity dominates the noise 
level at all separations, thus the calculated upper limit is likely 
conservative, overestimating the true residual speckle noise level.
We do not detect any statistically significant signals that could be 
considered companion candidates. 

\Citet{bonavita2010} present a more detailed discussion of 
point-source constraints around LkCa~15, including alternate
evolutionary model assumptions, based on the NACO $K_\mathrm{s}$-band
data.  They reach the same contrast levels as we do.

\begin{figure}[t]
\centering
\vspace*{2mm}
\includegraphics[width=0.95\linewidth]{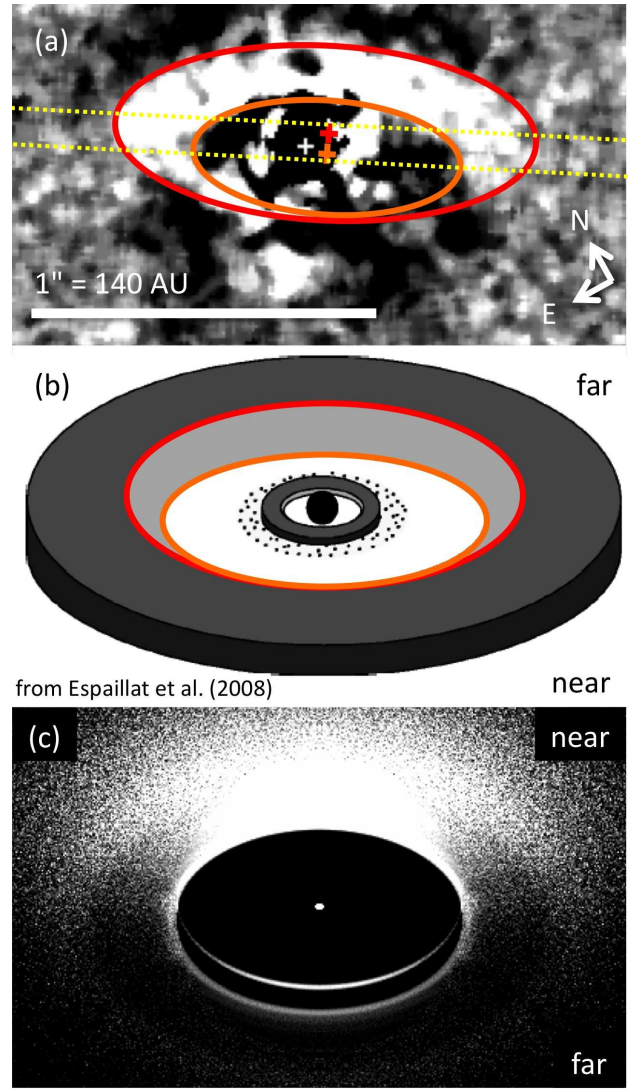}
\vspace*{1mm}
\caption{(a) Ellipse fits to the inner and outer boundaries of the 
	scattered light nebulosity seen in the HiCIAO 
	$H$-band LOCI image after median filtering on the
	spatial scale of 5 pixels $\approx$ 1\,FWHM and derotation by 
	$-29.3^\circ$
	\citep[based on the position angle of $150.7^\circ$ in][]{pietu2007}.
	The inner (orange)
	and outer (red) ellipses are offset from the star along the major axis 
	by 51\,mas and 57\,mas and rotated by $-4^\circ$ and $-3^\circ$,
	respectively.  Their centers are marked by orange and red plus signs, 
	respectively, while the star's position is indicated by a white plus
	sign. The dotted lines delimit the area on which the
	quantitative analysis in Figure~\ref{f:bisector} is based (offset
	angle $-3.5^\circ$).
	(b) Sketch of the illuminated wall scenario, 
	taken from \citet{espaillat2008}, based solely on 
	the SED and millimeter interferometry. The inner disk is not to 
	scale.  The illuminated disk wall on the far side (light grey) 
	is directly visible while the near-side wall blocks its bright side
	from view.
	(c) $H$-band image of the forward scattering scenario
	including anisotropic scattering, from the simulation presented in
	\citet{mulders2010}.
	The near-side disk surface (top) appears bright due to efficient 
	forward scattering, whereas the far-side disk wall (bottom) is mostly
	shadowed by the inner disk (center, not resolved), reducing the wall
	image to two thin parallel arcs. 
	}
\label{f:ellipses}
\end{figure}

We find a likely positional offset of the inner and outer boundaries
of the nebulosity from the star along the system semi-major axis 
(Figures~\ref{f:bisector}, \ref{f:ellipses}a).  Furthermore,
our inner and outer fitted ellipses are rotated  by $-4^\circ$ and 
$-3^\circ$ with respect to the position angle of $150.7^\circ$ 
in \citet{pietu2007}, with an 
estimated fitting error of $2^\circ$. 

In order to quantify the observed offset, we cut a 
strip with a width of 11\,pixels = 0\farcs10 centered on the star
and oriented along the major axis of our fitted ellipses from the LOCI 
image, assuming a $-3.5^\circ$ offset from the 1.4\,mm position 
angle.  Since the disk does not suffer from foreshortening along 
its apparent major axis, the distances along the strip are
a direct measure of physical distance.  We collapse the strip into
a one-dimensional profile, plotted in Figure~\ref{f:bisector}.

The intensity drops 
down to zero over a distance of 54\,mas ($\approx$ 8\,AU, 1.0\,FWHM)
on the West side and 81\,mas ($\approx$ 11\,AU, 1.5\,FWHM)
on the East side of the gap, indicating that the edge is sharp at or 
below the image resolution.  The bisecting curve
between the two opposed gap edges reveals a systematic offset from the
star's position by 64\,mas (9\,AU) with a standard deviation of 6\,mas 
(0.9\,AU).
Since the star is unsaturated, its centering accuracy ($\sim$0.2 pixels 
= 2\,mas) does not contribute significantly to the uncertainty.
The distance of the gap edges from the star are 345\,mas = 48\,AU on 
the left side and 447\,mas = 63\,AU on the right side at an accuracy of
9\,mas (1.3\,AU), consistent with the
46\,AU radius derived from millimeter interferometry by \citet{pietu2006}.

After scaling each pixel in the strip by $r^2$ before
collapsing the strip, we find that the nebulosity appears at roughly the
same brightness on both sides of the gap, suggesting a $r^{-2}$ dependency 
of the unscaled flux levels.  
This is consistent with the assumption that we are looking at reflected 
light from material at varying distances $r$ from the star.

Thus, the inner edge of the outer disk around LkCa~15
likely features a pericenter offset comparable to those 
observed in the disks of HD~142527 \citep{fukagawa2006} and Fomalhaut 
\citep{kalas2005}.

\section {Discussion} 


\subsection{Forward scattering scenario}

One possible explanation of the observed bright nebulosity is
forward-scattering, which is commonly invoked to explain the brightness
asymmetries in disk surfaces seen in reflected light 
\citep[e.g.][]{weinberger1999,fukagawa2006}.
Forward-scattering on large dust grains can be several
times as efficient as backward-scattering; as a result, the near-side 
surface of such a disk appears brighter than the far side.  This
requires the outer disk surface to be illuminated by the star.
The first of the two SED-compliant models in \citet{mulders2010} includes
a forward-scattering disk surface; furthermore, the far-side gap wall is
shadowed by an optically thick inner disk, rendering it hard to detect
in direct imaging (Figure~\ref{f:ellipses}c). 
Note that the size and shape of the
	dust grains in the wall are not well known.
This scenario is supported by \citet{pietu2007},
whose orientation and inclination values for the LkCa~15 disk suggest
that the northwest side is the near side, as well as the fact that
the crescent seen after reference PSF subtraction features a bright 
spot along the minor axis.

\subsection{Illuminated wall scenario}

Another explanation for the nebulosity is that it represents the inner
wall of the outer disk on the far side of the star, illuminated by the
star and viewed directly through the disk gap 
\citep[Figure~\ref{f:ellipses}b]{espaillat2008}.
The illuminated surface
of the near-side wall, on the other hand, is blocked from sight by the 
bulk of the optically thick disk. Furthermore, the wall is high enough 
to cast the outer surface of the disk into shadow, suppressing a 
forward-scattering signature.  This corresponds to the second model
by \citet{mulders2010}.
The morphology of the nebulosity favors this scenario, given that (1) 
there is a sharp outer edge roughly parallel to the inner edge, (2) the
nebulosity reaches across the major axis, ``embracing'' the star, and 
(3) the nebulosity is wider along the major axis than along the minor, 
all of which are expected for an illuminated tapered
disk wall, but not for forward scattering on the disk surface.  Thus,
we consider the illuminated wall scenario the more likely explanation
until further data become available.  

\subsection{Gap formation mechanism}

Both scenarios presented above must invoke a disk gap to explain the 
sharp inner edge of the observed nebulosity.  Our data therefore 
prove the validity of the gapped disk model predicted from the SED.

Several mechanisms have been proposed to produce cavities or gaps in
protoplanetary disks.
Magneto-rotational instability (MRI) as described in \citet{chiang2007} can
be excluded for LkCa~15, since it acts on all disk components that can
be ionized by direct X-ray illumination from the star, thus the inner
dust component of LkCa~15 known from the SED could not have survived.
Furthermore, the stellar mass of 0.97\,$M_\odot$ and stellar accretion 
rate of 2.4\,$\cdot$\,10$^{-9}$ $M_\odot$/yr \citep{espaillat2007} 
would require a viscosity parameter $\alpha\approx 0.0007$, one order
of magnitude below the range considered in the publication.
While photoevaporation can in principle evacuate gaps as large as that
of LkCa~15 in simulations \citep{alexander2006,gorti2009,owen2010}, 
those require that the inner disk drain away entirely before the gap can
grow beyond 1--10\,AU.  The presence of an inner dust disk around LkCa~15 
does not fit this scenario.

\Citet{espaillat2008} therefore conclude that dynamical clearing by
one or more orbiting bodies is the most plausible cause of the disk
gap.  Our detection of a likely pericenter offset in the LkCa~15 disk
gap reinforces this argument.  Dynamical sculpting is commonly 
invoked to explain warps and eccentricities in debris disks 
\citep[e.g.][]{roques1994,kalas2005}, and also produces 
off-centered gaps in simulations of protoplanetary disks
\citep[e.g.][]{marzari2010}.  In contrast, neither MRI nor 
photoevaporation have been predicted to cause such effects.

In the case of LkCa~15, close stellar companions are excluded by 
long-baseline interferometry \citep{pott2010}, while our data 
conservatively exclude companions more massive than 21\,$M_\mathrm{Jup}$ 
exterior to 0\farcs1 (14\,AU),  and  all but planetary-mass bodies 
exterior to 0\farcs2 (28\,AU). 
\Citet{pietu2006} note that a 5--10\,$M_\mathrm{Jup}$ body orbiting at 
30\,AU (the equivalent of Neptune's orbit) has a sufficiently large Hill 
sphere to dynamically produce a wall at 50\,AU.  Less massive bodies 
orbiting closer to the wall can have similar dynamical effects.
\Citet{lubow1999} note that bodies more massive than 6\,$M_\mathrm{Jup}$ 
will 
suppress accretion onto a young Solar analog to levels below that still 
present in the LkCa~15 system.  Our data indicate that any
bodies in the disk of LkCa~15 at $r\ge 0\farcs2$ (28\,AU) from the 
primary in projection must have planetary masses.  

We therefore find dynamical clearing by
one or more planets to be the most likely cause of LkCa~15's disk
gap. Given the Sun-like mass of
the young star ($0.97\,M_\odot$), we might in fact be looking at a Solar
System analog in the making.


\acknowledgements
We thank Cornelis P.\ Dullemond and Dmitry Semenov for helpful discussion,
and David Lafreni\`ere for generously providing us with the
source code for his LOCI algorithm.
This work is partly supported
by a Grant-in-Aid for Science Research in a Priority Area from MEXT and
by the Mitsubishi Foundation.  JPW and MWM acknowledge support from NSF 
Astronomy \& Astrophysics Postdoctoral Fellowships AST 08-02230 and 
AST-0901967, respectively.  JPW also acknowledges funding from a 
Chr\'etien International Research Grant. ELT gratefully acknowledges 
support from a 
Princeton University Global Collaborative Research Fund grant and the 
World Premier International Research Center Initiative (WPI Initiative),
MEXT, Japan.  Part of this work was carried out at the Jet Propulsion 
Laboratory, California Institute of Technology, under contract with NASA.

{\it Facilities:} \facility{Subaru (HiCIAO, AO188)}, 
\facility{VLT (NACO)}.

\clearpage

\end{document}